\documentstyle[12pt]{article}
\author{M.P. Pato, C.A. Nunes, C.L. Lima, M.S. Hussein\\
Nuclear Theory and Elementary Particle Phenomenology Group\\
Instituto de F\'\i sica, Universidade de S\~ao Paulo,\\
Caixa Postal 20516, 01498-970, S\~ao Paulo, SP, Brasil
\and
Y. Alhassid\\
Center for Theoretical Physics, Sloane Physics Laboratory\\
Yale University, New Haven, CT06511, USA}
\title{Deformed Gaussian Orthogonal Ensemble Analysis of the Interacting Boson
Model}
\begin{document}
\maketitle
\begin{abstract}
A Deformed Gaussian Orthogonal Ensemble (DGOE) which interpolates between
the Gaussian Orthogonal Ensemble and a Poissonian Ensemble is constructed.
This new ensemble is then applied to the analysis of the chaotic properties
of the low lying collective states of nuclei described by the Interacting
Boson Model (IBM). This model undergoes a transition order-chaos-order from
the $SU(3)$ limit to the $O(6)$ limit. Our analysis shows that the quantum
fluctuations of the IBM Hamiltonian, both of the spectrum and the
eigenvectors, follow the expected behaviour predicted by the DGOE when one
goes from one limit to the other. \thinspace
\end{abstract}

It is widely assumed that Random Matrix Theories (RMT) \cite{bohigas}\
provide a basis to study quantum chaotic systems. In particular, it is
expected that fluctuation properties of fully chaotic systems with time
reversal symmetry follow the Gaussian Orthogonal Ensemble (GOE) whereas
non-chaotic ones follow the Poissonian Ensemble. Some physical systems
however may exhibit statistics intermediate between these two limits, as
recent investigations have shown in the case of the excitation spectra and
intensities of deformed \cite{alhassid}\cite{broglia} and spherical \cite
{otsuka} nuclei. The analysis that has been performed on these systems have
a more or less empirical character in the sense that they are not based
directly on an ensemble of RMT. It is therefore important to test the
reliability of the RMT predictions in these intermediate situations.

Recently, two of us have constructed a Deformed Gaussian Orthogonal Ensemble
(DGOE) using the maximum entropy principle applied to generic random
matrices subjected to appropriate constraints\cite{pato1}. The constraints
imposed were such that the ensemble obtained goes from a pure GOE to a
combination of two GOE's. This corresponds to the case of $SU(2)$ symmetry
breaking, when a quantum number which can have only two values, e.g.
isospin, is partially conserved\cite{pato2}. To deal with the more general
case of the GOE-Poisson transition an extension of the DGOE of Ref.\cite
{pato1} is required. In this paper we provide this extended DGOE and apply
it to the analysis of the transition chaos-order exhibited by the
Interacting Boson Model (IBM) Hamiltonian.

We start using projection operators $P_i=\mid i><i\mid $, $Q_i=1-P_i$, where
$\mid i>$, with $i=1,2,...,N$, are N abstract basis vectors, to split a
generic Hamiltonian operator H in its diagonal and off-diagonal parts, $H_0$
and $H_1$. Namely,

\begin{equation}
\label{01}H_0 = \sum_{i=1}^N \; P_i\, H\, P_i \;\;,
\end{equation}

\begin{equation}
\label{02}H_1 = \sum_{i=1}^N \; P_i\, H\, Q_i
\end{equation}

\noindent where $H_0$ and $H_1$ satisfy the identity $H=H_0+H_1$.

Following the discussion of Ref.\cite{pato1} we add to the usual GOE
constraints

\begin{equation}
\label{04}< Tr H^2 > = \int\; dH\, P(H)\, Tr\, H^2 = \mu
\end{equation}

\begin{equation}
\label{05}< 1 > = \int\; dH\, P(H) = 1
\end{equation}

\noindent
the additional one

\begin{equation}
\label{06}<Tr\,H_1^2>=\sum_{i=1}^N\;\int
\;dH\,P(H)\,Tr(P_i\,H\,Q_i\,H\,P_i)=\nu .
\end{equation}

Maximizing then the entropy subjected to the above conditions we get, after
normalization, the probability distribution:

\begin{equation}
\label{08}P(H)=P_{GOE}(H)exp[-\beta \sum_iTr(P_iHQ_iHP_i)](1+\frac \beta
\alpha )^{\frac N4(N-1)}
\end{equation}

\noindent where $\alpha $ and $\beta $ are Lagrange multipliers and

\begin{equation}
\label{09}P_{GOE}(H)=2^{-N/2}\left( \frac \pi {2\alpha }\right) ^{-\frac
14N(N-1)}exp(-\alpha \,Tr\,H^2)
\end{equation}

\noindent is the GOE distribution.

It is clear from Eq. \ref{08} that when $\beta \rightarrow 0$ we recover the
GOE distribution while with the increase of $\beta $ the system becomes less
chaotic; in the limit $\beta $ $\rightarrow \infty $ we recover the
Poissonian Ensemble .

Using an idea suggested by Dyson \cite{dyson}, Alhassid and Levine\cite
{levine} discussed an intermediate ensemble as a solution of a
non-equilibrium problem defined by an appropriate stochastic Langevin
equation. The non-integrability is characterized by a parameter $\epsilon $
which is the ratio between the variances of the non-diagonal to the diagonal
elements of H. We easily find the relation $\epsilon =\frac 1{\sqrt{1+\beta
/\alpha }}$ showing that $\epsilon $ goes from zero to unity as $\beta $
varies from infinity to zero.

At this point, we remark that by using the projectors $P_i$ and $Q_i$ we can
alternatively decompose the Hamiltonian $H$ as

\begin{equation}
\label{13}H=H_0+\epsilon \sum_{i=1}^N\;P_i\,H_G\,Q_i
\end{equation}

\noindent
where $H_G=H(\beta =0)$ showing that the problem may be reformulated in such
a way that the parameter $\epsilon $ appears as a coupling constant. This
relation makes a connection of our ensemble to the one discussed recently by
Lenz and Haake\cite{lenz}.

We present now the numerical results. In Ref.\cite{levine} a new parameter $%
\tilde \epsilon $ was introduced such that the GOE limit is expected to be
approached when $\tilde \epsilon \rightarrow 1$. This parameter may be
expressed as

\begin{equation}
\label{14}\tilde{\epsilon} = \sqrt{\frac{2\nu}{\mu - \nu}} = \epsilon \sqrt{%
N-1}
\end{equation}

\noindent and for N=2 $\tilde \epsilon =\epsilon $ while for large N, $%
\tilde \epsilon \sim \sqrt{N}\epsilon $. To test this idea we have plotted
in Fig.1 the parameter $\omega $ of the Brody distribution that fits the
level spacing as a function of the size N of the matrix, keeping $\epsilon $
or $\tilde \epsilon $ fixed. We fixed the value of $\tilde \epsilon $ to be
that at N=50, the lowest dimension in the calculation. It is clear that $%
\omega $ saturates quite rapidly when $\tilde \epsilon $ is kept constant.

In Fig.2, we have considered matrices of size N=200 and calculated the level
spacing P(s), the spectral rigidity $\Delta _3(L)$ and the distribution P($%
\ln y$), where $y$ is the square of the component normalized with respect to
its average $(y=c^2/<c^2>)$, for several values of the parameter $\tilde
\epsilon $. We see that indeed we are very close to GOE when $\tilde
\epsilon =1$. The level spacing distribution exhibits the universality law
which says that the level repulsion only disappears in the Poisson limit, $%
\tilde \epsilon $ = 0 or $\beta =\infty $. We remark also that our P(s) are
practically identical to those obtained by Lenz and Haake with N = 500
matrices. With respect to P($\ln y$), the distributions get broader as $%
\tilde \epsilon $ decreases but we cannot fit them with only one $P_\nu .$
Actually, as in the previous case of the DGOE for two GOE's, an excellent
fit is obtained if we use instead two $P_\nu $'s. This behavior may be
understood as a consequence of the fact that in the transition from the GOE
to the Poissonian case, the components of the eigenvectors go from a
situation in which they are equally distributed among the basis states, to a
limit situation in which only one intensity, say $y_i$, is different from
zero. Near this limit, the average of the components becomes equal to $%
<y>\simeq 1-y_i$ and this explains why the distribution shift to the left in
Fig.2 . This fact, together with the splitting of the intensities in two
sets fitted by two different $P_\nu $'s, seem to be universal features of
the eigenvector statistics in the transition from a chaotic to a regular
regime.

The Interacting Boson Model (IBM) has been successfully applied to the
phenomenological description of low-lying collective states of atomic nuclei%
\cite{iachello}. In Ref.\cite{alhassid} the fluctuation properties of such
states were analyzed in the framework of IBM.We turn now to a detailed
analysis of the statistical properties of the IBM Hamiltonian, using the
above intermediate ensemble. The IBM Hamiltonian may be written as

\begin{equation}
\label{15}H=\eta \,n_d-(1-\eta )Q^\chi \cdot Q^\chi
\end{equation}

\noindent
where $n_d$ is the number of $d$ bosons and $Q^\chi $ is the quadrupole
operator

\begin{equation}
\label{16}n_d=d^{\dagger }\cdot \tilde d
\end{equation}

\begin{equation}
\label{17}Q^\chi =(d^{\dagger }\times s+s^{\dagger }\times \tilde
d)^{(2)}+\chi (d^{\dagger }\times \tilde d)^{(2)},
\end{equation}

with the relation $\tilde d_\mu =(-)^\mu d_{-\mu }$. The six bosons $s$ and $%
d_\mu $ span a six-dimensional Hilbert space which has $U(6)$ symmetry. The
IBM Hamiltonian has three dynamical symmetry limits which describe
vibrational nuclei for $\eta =1$, rotational nuclei for $\eta =0$ and $\chi
=-\sqrt{7}/2$, and, finally, $\gamma $-unstable nuclei for $\eta =0$ and $%
\chi =0$. It was shown recently \cite{alhassid} that among these three
limiting cases, there is a region where the Hamiltonian becomes chaotic. In
particular, taking $\eta =0$ and various values of $\chi $ in the range $-
\sqrt{7}/2\leq \chi \leq 0$, namely going from rotational nuclei to $\gamma $%
-unstable ones we encounter the following results shown in Fig.3. With
regards to the level spacing and the $\Delta _3$ we are just reproducing the
results of Ref.\cite{alhassid}. However, instead of presenting the BE(2)
distributions as was done in Ref.\cite{alhassid}, we show in the first
column of the figure the statistics of the components. This is necessary in
order to compare with the DGOE statistics presented in Fig.2. Besides, the
components were calculated with respect to the basis of the nearest regular
cases.

The IBM spacings and $\Delta _3$ distributions show a clear
Poisson-GOE-Poisson transition as one varies $\chi $ from -1.25 to -0.1. The
two transitions considered separately have a behavior which compares well
with the description given by the ensemble. However, insofar as the
component distributions are concerned, there is a marked difference, which
seems to arise from what we may call the intermediate structure that
modulates the components distribution. This intermediate structure was
refered to as secular variations in Ref.\cite{alhassid}. Therefore in order
to make a sensible comparison we still have to extract from the components
this secular variation. This can be performed by introducing a local average
defined by an appropriate moving window. Following the procedure of Ref.\cite
{alhassid} we have used a Gaussian window and have taken the labels of the
components as variables. Explicitly, we have used

\begin{equation}
\label{20}\bar y(a,b)=\frac{\sum_{i=1}^N\sum_{j=1}^N\left( c_i^j\right)
^2\exp \left( -\frac{\left( a-i\right) ^2}{2\gamma ^2}\right) \exp \left( -
\frac{\left( b-j\right) ^2}{2\gamma ^2}\right) }{\sum_{i=1}^N\sum_{j=1}^N%
\exp \left( -\frac{\left( a-i\right) ^2}{2\gamma ^2}\right) \exp \left( -
\frac{\left( b-j\right) ^2}{2\gamma ^2}\right) }
\end{equation}

and the normalized components are defined as $y_a^b=\frac{\left(
c_a^b\right) ^2}{\bar y(a,b)}$ . In fig.4 we show the resulting
distributions. In the column on the left the we have the distributions for
the DGOE in the case of 200$\times $200 matrices whereas the other two
present the IBM results. The remaining two columns show the transitions
order-chaos starting near the rotational limit ($\chi =-1.25$, bottom) and
starting near the $\gamma -$unstable limit ($\chi =-0.10$, bottom),
respectively. We can see that the distributions for the two transitions of
the IBM Hamiltonian follow the pattern described by the ensemble.

In conclusion we have shown in this paper that the extended version of the
DGOE of Ref.\cite{pato1}, can indeed describe well the statistical behavior
of a realistic nuclear model such as the IBM. Similar observations can be
made with regard to the study of high-spin states \cite{broglia}, and the
analysis of spherical nuclei made in Ref.\cite{otsuka}. We have reasons to
believe that any physical system is said to exhibit order-chaos transition
when its statistics follows rigorously that of a extended DGOE.

\newpage\ {\bf Figure Captions:}

Fig.1-The Brody parameter, $\omega $, of the adjusted spacing distributions,
vs. the size of the matrix. The diamonds correspond to $\epsilon =0.01$,
whereas the crosses to $\epsilon =\tilde \epsilon /\sqrt{N-1}=0.07/\sqrt{N-1}
$.

Fig.2-The distributions, P($\ln y$), P($s$), and $\Delta _3$ of the DGOE for
N=200 and various values of $\epsilon $. See text for details.

Fig.3-Same as Fig.2, for J=$6^{+}$ IBM states. See text and Ref.\cite
{alhassid} for details.

Fig.4-The components distributions after normalization, the lhs column
presents the DGOE 's and the other two the IBM ones . See text for details.

\end{document}